\documentclass[aps,showpacs,twocolumn,preprintnumbers,amsmath,amssymb,groupedaddress]{revtex4}
\usepackage{graphicx}
\begin{document}
\title{Elastic properties of FeSi}

\author{Alla E. Petrova, Vladimir N. Krasnorussky}
\author{ Sergei M. Stishov}  \email{sergei@hppi.troitsk.ru}

\affiliation{Institute for High
Pressure Physics of Russian Academy of Sciences, Troitsk, 142190 Moscow
Region, Russia}

\date{\today}

\begin{abstract}
Measurements of the sound velocities in a single crystal of FeSi were performed in the temperature range 4-300 K. Elastic constants $C_{12}$ and $C_{44}$ deviate from a quasiharmonic behavior at high temperature; whereas, $C_{12}$ increases anomalously in the entire range of temperature, indicating a change in the electron structure of this material.
\end{abstract}


\maketitle
The intermetallic compound with a cubic B20 crystal structure FeSi has attracted much attention over decades due to its unusual physical properties. FeSi, being a semiconductor with a small energy gap of about 0.05 eV at low temperature, reveals metallic properties above 100 K~\cite{1,2,3,4,5}. Angle resolved photoemission spectroscopy demonstrates the disappearance of an energy gap in FeSi at high temperatures~\cite{6,7,8}. One of the intriguing features of the physical properties of  FeSi is the growth of its magnetic susceptibility at temperatures above 100 K, which reaches a broad maximum at 500 K, followed by Curie Weiss behavior at higher temperatures. Its electrical conductivity also rises steeply in the same temperature interval as the magnetic susceptibility. No magnetic ordering in FeSi was found at any temperature~\cite{9}. A number of theories and models have been suggested to explain the experimental observations, including a model of localized states at the Fermi level, a spin-fluctuation theory and a Kondo model (see for instance~\cite{1,3,10}). According to recent studies, the Kondo scenario is not needed for an explanation of the physical properties of FeSi~\cite{8}. The concept of a strongly correlated semiconductor with an energy gap destroying by the correlations has gained acceptance as a general framework for interpreting properties of FeSi. But, independently from any kind of theories, experimental data show that an insulator-metal crossover takes place in FeSi at elevated temperature, which should influence the inter-particle interactions and hence the elastic properties of FeSi

The main goal of the present paper is to study effects of the gradual metallization on the elastic properties of FeSi.  The significance of this study is enhanced by the importance of FeSi as a potential constituent of the Earth's core. Earlier, elastic properties of FeSi were studied in Ref. ~\cite{11} and ~\cite{12} in the temperature range $\sim$80-400 K, but, as we find, extending these measurements to lower temperatures provides a more complete interpretation of the origin of its elastic properties.

We report here results of ultrasonic studies of a single crystal of FeSi in the temperature range 4-300K. The measurements were performed using a digital pulse-echo technique (see details in Ref.~\cite{13}). The single crystal of FeSi was grown by the Chokhralsky method. The lattice parameter of the crystal, determined by powder X-ray diffraction, was equal to 4.483 {\AA}, which corresponds well to literature values. Electron-probe microanalysis showed practically a stoichiometric chemical composition, which was confirmed by an X-ray Rietveld refinement of the diffraction pattern.

\begin{figure}[htb]
\includegraphics[width=80mm]{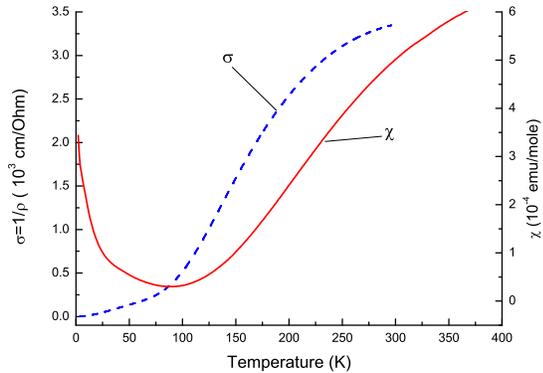}
\caption{\label{fig1} (Color online)  The  temperature dependence of dc conductivity ($\sigma$) and magnetic susceptibility ($\chi$) of FeSi.}
\end{figure}

\begin{figure}[htb]
\includegraphics[width=80mm]{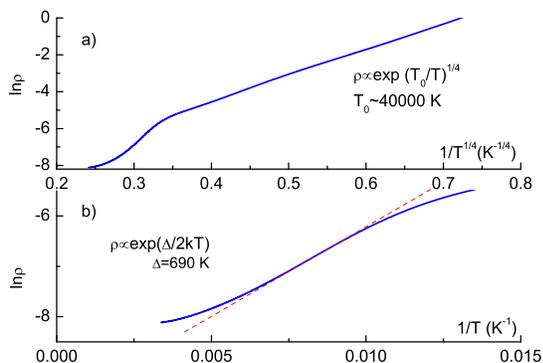}
\caption{\label{fig2} (Color online) Electrical conductivity $\sigma$ on a logarithmic scale as a function of  $T^{-1/4}$(a) and $T^{-1}$(b) for FeSi. }
\end{figure}

Electrical and magnetic characteristics of the crystal are displayed in Figs.~\ref{fig1} and~\ref{fig2}. Electrical measurements were made by a four terminal dc method. Magnetic susceptibility was measured with a Quantum Design Magnetic Properties Measurement System.  Fig.~\ref{fig2} shows the temperature dependence of conductivity and magnetic susceptibility of the FeSi samples, cut from the same batch as those used for ultrasound studies.  Note that the steep increase in both the conductivity and magnetic susceptibility start at about 100 K.  Fig.~\ref{fig2}a demonstrates that in the temperature regime below about 100 K the resistivity can be approximated by the expression for variable range hopping. In the narrow temperature region of 100-150 K, the resistivity can be described by a standard activation formula with an energy gap Eg of 0.06 eV or 690 K (Fig.~\ref{fig2}b). At higher temperatures the resistivity crosses over into a saturation regime, therefore indicating the gap closing and entry into a metallic state. All these observations generally agree with literature data (see for instance~\cite{5}).

For the ultrasonic studies, two samples of FeSi of about 2 mm thickness and with orientations along [111] and [100] were cut from a big single crystal. The corresponding surfaces of the samples were made optically flat and parallel. The $36^{o}$ Y (P-wave) and $41^{o}$ X (S-wave) cut $LiNbO_3$ transducers were bonded to the samples with various adhesives, including silicon greases and superglue. Temperature was measured by a calibrated Cernox sensor with an accuracy of 0.02 K.

\begin{figure}[htb]
\includegraphics[width=80mm]{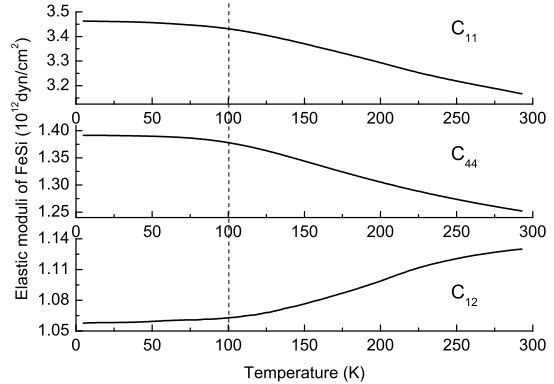}
\caption{\label{fig3} The temperature dependence of elastic constants of FeSi}
\end{figure}

\begin{figure}[htb]
\includegraphics[width=80mm]{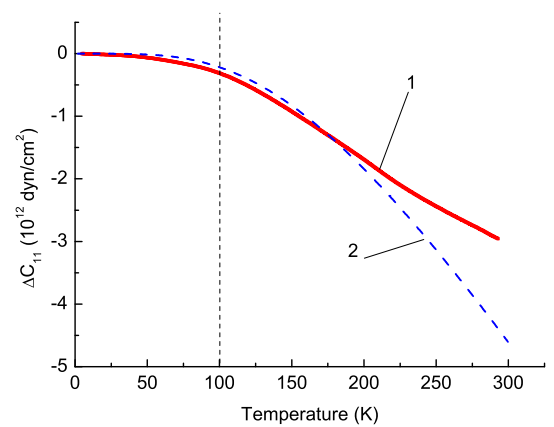}
\caption{\label{fig4} (Color online) $C_{11}= C_{11}(T=0) - C_{11}(T)$ as a function of temperature. 1-experimental data, 2-extrapolation from low temperatures with the formula~\ref{eq:1}.}
\end{figure}

\begin{figure}[htb]
\includegraphics[width=80mm]{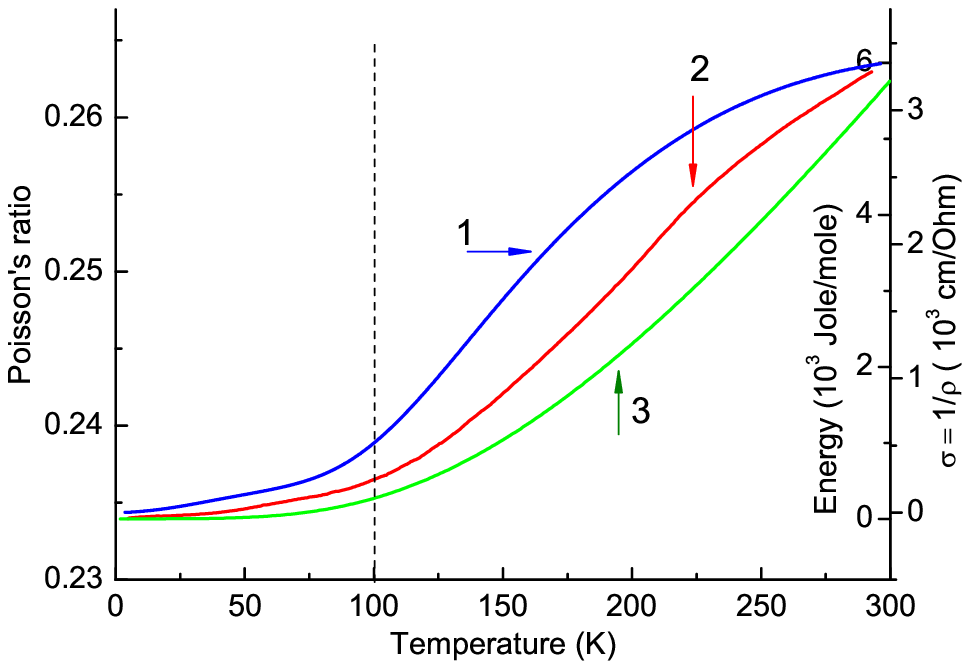}
\caption{\label{fig5} (Color online) The conductivity 1, Poisson's ratio 2 and thermal energy 3 for FeSi versus temperature. The thermal energy is calculated in the Debye approximation ($\Theta_D$=690 K)}
\end{figure}

The speed of sound and elastic constants are calculated using the known thickness and density of the samples and the relationship $C_{ij} = \rho V^2$.  The absolute accuracy of the sound velocity measurements is about 0.1\%, mainly due to uncertainty connected with a phase shift at the transducer-sample bond interface, but the accuracy of some elastic constants may be of order 1\%,  if calculated using data from different runs with samples of different orientations.

\begin{table}
\caption{\label{tab:table1} Elastic moduli of FeSi ($10^{12}$ $dyn/cm^2$), a = 4.483 ${\AA}$, 298 K }
\begin{ruledtabular}
\begin{tabular}{cccccccc}
$C_{ij}$&$T=4.8 K$&$T=77.8 K$&$T=292.8 K$\\
\hline
$C_{11}$& 3.462 & 3.445 & 3.167 \\
$C_{44}$& 1.390 & 1.386 & 1.252 \\
$C_{12}$& 1.058 & 1.060 & 1.129 \\
\end{tabular}
\end{ruledtabular}
\end{table}

Results of the measurements and calculations are shown in Fig.~\ref{fig3} and Table~\ref{tab:table1}. Our data on FeSi generally agree with results of Refs.~\cite{11,12} in the overlapping region of temperatures, though relatively lower resolution in Ref.~\cite{11} does not allow a detailed comparison.

As is seen from Fig.~\ref{fig3}, the elastic constants $C_{11}$ and $C_{44}$ of FeSi decrease with temperature at a rate growing quickly at about 100 K. This behavior can be explained most straightforwardly by a simple model of solids. According to quasiharmonic theory, a temperature dependence of the adiabatic elastic constants is defined by the expression~\cite{14}
\begin{equation}
\label{eq:1}
C_{ij}=C_{ij}(1-D\bar{\epsilon})
\end{equation}
where $C_{ij}$ is harmonic values of elastic constants, $D$ is a parameter, defined by the crystal structure, $\bar{\epsilon}$ is the mean energy per oscillator ( $3sN\bar{\epsilon}=U$, where $U$ is thermal energy, $N$ is the number of atoms, and $s$ is the degree of freedom).

Using values of the elastic constants from Table~\ref{tab:table1}, we estimate the Debye temperature of FeSi, which appeared to be $\Theta_D$=690 K, and calculated the corresponding thermal energy $U$.  Simple calculations, illustrated in Fig.~\ref{fig4}, show that the Debye model of solids can account for the behavior of $C_{11}$ and $C_{44}$ to at least 170K. At the same time, these calculations reveal anomalous stiffening of $C_{11}$ and $C_{44}$  at higher temperatures (see in this connection Ref.~\cite{12}. The growth of $C_{12}$ (Fig.~\ref{fig3}) and hence a steady increase of the Poison's ratio $C_{12}/(C_{12}+C_{11})$ (Fig.~\ref{fig5}) in the entire range of temperatures is another specific feature of the elastic properties of FeSi. The latter was not properly recognized in Ref.~\cite{12}, though an anomalous behavior of $C_{12}$ can be seen after a careful analysis of the figures in~\cite{12}.

An explanation of that highly "anharmonic" behavior of the elastic constants of FeSi apparently lies in the specifics of its electronic subsystem. The emergence of a metallic state implies generation of an  electron liquid  in FeSi that normally would give a positive contribution to its bulk modulus and modify the interatomic interaction. Both factors are expected to inluence the elastic constants, which certainly could lead to a deviation from a simple quasiharmonic model.

Remarkably, the energy gap and the Debye temperature $\Theta_D$ in FeSi have practically the same values. This fact is reflected in a fast growth of the charge carrier and phonon populations, obviously occurring in the same temperature interval around 100 K, as it follows from behavior of the electrical resistivity, magnetic susceptibility and elastic constants of FeSi. It is worth recalling that the mechanism of conductivity in FeSi changes from hopping to activation also around 100 K (Fig.~\ref{fig2}). However, whether this is a simple coincidence or whether there is hidden physics in the background remains to be seen.

Finalizing, we note that anharmonic behavior of elastic constants  suggests changes in the character of interatomic interactions in FeSi, as is evidenced by the metal - insulator crossover. Probably anomalous behavior of $C_{12}$ and the Poisson's ratio could serve as indicators of a qualitative change in the electronic structure of materials. Indeed, Fig.~\ref{fig5} explicitly shows the generic connection between variations of conductivity, the Poisson's ratio and thermal energy in FeSi.

Authors express their gratitude to A.A. Mukhin, V.Y. Ivanov, I.P. Zibrov, N.F. Borovikov and V.O. Yapaskurt for help in characterization of the sample, which was kindly provided by A.C. Ivanov. J.D. Thompson read the paper and made valuable comments. We appreciate support of the Russian Foundation for Basic Research, Program of the Physics Department of RAS on Strongly Correlated Systems and Program of the Presidium of RAS on Physics of Strongly Compressed Matter.

\end{document}